\documentclass[
 preprint,
 superscriptaddress,
 amsmath,amssymb,
 aps,
 prl,
 longbibliography,
]{revtex4-2}

\usepackage{graphicx}
\usepackage{dcolumn}
\usepackage{bm}
\usepackage{natbib}
\usepackage{hyperref}

\begin{document}


\title{Chiral photonic circuits for deterministic spin transfer}

\author{Shan Xiao}

\author{Shiyao Wu}

\author{Xin Xie}
\author{Jingnan Yang}
\affiliation{Beijing National Laboratory for Condensed Matter Physics, Institute of Physics, Chinese Academy of Science, Beijing 100190, China}
\affiliation{CAS Center for Excellence in Topological Quantum Computation and School of Physical Sciences, University of Chinese Academy of Sciences, Beijing 100049, China}

\author{Wenqi Wei}

\author{Shushu Shi}
\author{Feilong Song}
\author{Sibai Sun}
\author{Jianchen Dang}
\author{Longlong Yang}
\affiliation{Beijing National Laboratory for Condensed Matter Physics, Institute of Physics, Chinese Academy of Science, Beijing 100190, China}
\affiliation{CAS Center for Excellence in Topological Quantum Computation and School of Physical Sciences, University of Chinese Academy of Sciences, Beijing 100049, China}

\author{Yunuan Wang}
\affiliation{Beijing National Laboratory for Condensed Matter Physics, Institute of Physics, Chinese Academy of Science, Beijing 100190, China}
\affiliation{Key Laboratory of Luminescence and Optical Information, Ministry of Education, Beijing Jiaotong University, Beijing 100044, China}

\author{Sai Yan}
\author{Zhanchun Zuo}
\affiliation{Beijing National Laboratory for Condensed Matter Physics, Institute of Physics, Chinese Academy of Science, Beijing 100190, China}
\affiliation{CAS Center for Excellence in Topological Quantum Computation and School of Physical Sciences, University of Chinese Academy of Sciences, Beijing 100049, China}

\author{Ting Wang}
\author{Jianjun Zhang}
\author{Kuijuan Jin}

\author{Xiulai Xu}
\email{xlxu@iphy.ac.cn}
\affiliation{Beijing National Laboratory for Condensed Matter Physics, Institute of Physics, Chinese Academy of Science, Beijing 100190, China}
\affiliation{CAS Center for Excellence in Topological Quantum Computation and School of Physical Sciences, University of Chinese Academy of Sciences, Beijing 100049, China}
\affiliation{Songshan Lake Materials Laboratory, Dongguan, Guangdong 523808, China}

\date{\today}

\begin{abstract}

Chiral quantum optics has attracted considerable interest in the field of quantum information science. Exploiting the spin-polarization properties of quantum emitters and engineering rational photonic nanostructures has made it possible to transform information from spin to path encoding. Here, compact chiral photonic circuits with deterministic circularly polarized chiral routing and beamsplitting are demonstrated using two laterally adjacent waveguides coupled with quantum dots. Chiral routing arises from the electromagnetic field chirality in waveguide, and beamsplitting is obtained via the evanescent field coupling. The spin- and position-dependent directional spontaneous emission are achieved by spatially selective micro-photoluminescence measurements, with a chiral contrast of up to 0.84 in the chiral photonic circuits. This makes a significant advancement for broadening the application scenarios of chiral quantum optics and developing scalable quantum photonic networks.

\end{abstract}
\maketitle

\section{\label{sec1}Introduction}

Optical quantum systems have been at the research forefront among various quantum systems \cite{OBrien2009}. Photons have the advantages of a long coherence time, long-distance propagation, and ease of manipulation and detection, which show great potential for numerous applications in quantum technologies \cite{oBrien2007,kok2007,jin2010experimental,aspuru2012,spring2013,wang2017high,wang2018monolithic,wang2019,zhang2019integrated,kim2020}. For example, in quantum photonic integrated circuits (PICs), flying photons can connect stationary quantum nodes to achieve an efficient on-chip quantum interface, facilitating the construction of a scalable quantum network\cite{cirac1997quantum,ritter2012elementary,kalb2017entanglement,grim2019scalable}. Generally, spin states can be considered as matter qubits and used for quantum nodes for processing and storage operations. Such states can be mapped to circularly polarized photons generated by quantum emitters (e.g., quantum dots (QDs)) during an optical transition \cite{nakaoka2004size,atature2006quantum,gerardot2008optical,xu2008plug,vamivakas2009spin,warburton2013,lodahl2015,aharonovich2016,keil2017solid,tang2018,wu2020electron}. To obtain a reliable light-matter quantum interface, the deterministic coupling of quantum emitters and photons is essential. A promising approach is to exploit chiral quantum optics to form a chiral interface that facilitates the unidirectional transfer of the spin to the guided photons \cite{lodahl2017}. 

To date, chiral coupling has been intensively evaluated in various nanophotonic structures including metal surfaces \cite{rodriguez2013near,bliokh2015quantum}, optical fibers \cite{petersen2014chiral,mitsch2014quantum}, semiconductor waveguides \cite{luxmoore2013,sollner2015,coles2016,mrowinski2019}, microresonators \cite{junge2013strong,shomroni2014all,rodriguez2014resolving,cao2017experimental,liu2018transporting,martin2019chiral,tang2019chip,Yang2020}, and topological nanostructures \cite{barik2018,para2020}. Particularly, the tightly confined light field carries transverse spin angular momentum, thus a link between the spin and propagation direction of light can be introduced. Consequently, spin-momentum-locked light coupled to quantum emitters with circularly polarized transitions becomes directional-dependent \cite{abujetas2020}. However, previously reported chiral interfaces are relatively simple in  terms of functionality. The versatility of chiral interfaces is yet to be explored, such as the on-chip beam splitter, one of the main building blocks of PICs. Recently, arbitrary proportional beamsplitting at the single-photon level has been demonstrated experimentally in an active beam splitter \cite{prtljaga2014,jons2015,bishop2018,schnauber2018deterministic,papon2019nanomechanical}. However, engineering \textit{chiral} effects at the single-photon level in photonic circuits is highly desired and significantly more challenging.

Here, we demonstrate compact chiral photonic circuits using QDs as spin-polarized single-photon sources for deterministic spin transfer. First, we show that highly directional emission arises from the electromagnetic field chirality of a waveguide. Harnessing the chiral effect, we then numerically design a chiral beam splitter consisting of two laterally adjacent waveguides to form chiral photonic circuits. Circularly polarized chiral routing and beamsplitting are experimentally demonstrated in the chiral photonic circuits, with a chiral contrast of up to 0.84. The chiral routing direction can also be controlled by changing the QD position. The demonstrated circularly polarized chiral routing and beamsplitting are implemented in a conceptually different way, without utilizing intrinsic material properties and chiral structures \cite{konishi2011circularly,turner2013miniature,khorasaninejad2014silicon,fang2019ultra}, which offers the potential for developing on-chip functionalized optical elements and chiral quantum networks.

\section{\label{sec2}Results and Discussion}

\begin{figure}
\centering
\includegraphics[width=0.7\textwidth]{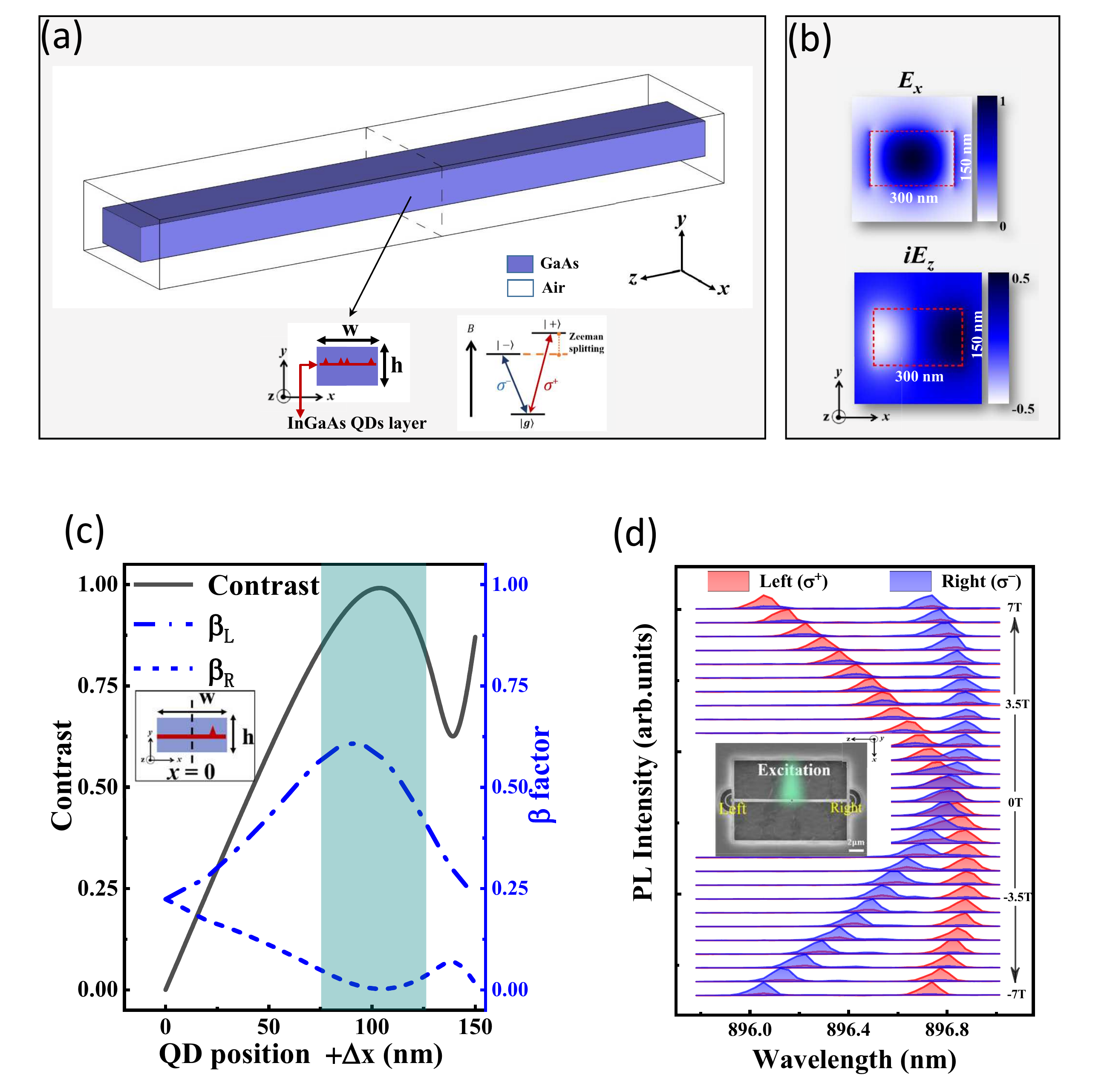}
\caption{Chirality in a suspended strip waveguide. a) Schematic view of a GaAs strip waveguide with full air cladding. Insets: cross-sectional view ($z=0$) of the waveguide with InGaAs QDs embedded in the center of the GaAs membrane and QD-level structure under magnetic field in the growth direction. b) Calculated fundamental TE-polarized electric field components. The upper panel is the real part of $ E_x $ whereas the lower panel is the imaginary part of $ E_z $. c) The directional $ \beta $ factor and chiral contrast as a function of the QD position $ +\Delta x $. $ +\Delta x $ represents the dipole displacement along the transverse direction $+x$ from the waveguide center to the edge, as shown in the inset. The cyan area indicates the high contrast exceeding 0.85 with QD approximately located off-center between approximately 75 nm and 125 nm. The calculation has been focused on the fundamental mode of an infinite waveguide, with a cross section of $300$ nm$\times$150 nm and a wavelength of 900 nm. The parameters used in the waveguide simulation remain consistent throughout the paper. All calculations are performed using the finite-difference time-domain method. d) PL spectra of a QD collected from left (red filling lines) and right (blue filling lines) OCs with a magnetic field from -7 T to 7 T. Inset: SEM image of the waveguide.}
\label{F1}
\end{figure}

\subsection{Chirality in a suspended strip waveguide}
Figure \ref{F1}(a) schematically shows a GaAs strip waveguide with full air cladding. A single layer of InGaAs QDs is embedded at the center of the GaAs membrane. According to Maxwell's equations, when ignoring the change of the space profile in the $ \pm z $ direction of propagation, the equation relevant to the electric field component can be expressed as $ {\varepsilon _{ \pm ,z}} =  \mp i/k(\partial {\varepsilon _{ \pm ,{\rm{x}}}}/\partial x + \partial {\varepsilon _{ \pm ,y}}/\partial y) $. The waveguide mode possesses a longitudinal field component and its strength depends on the transverse confinement. The proportionality fator $ \mp i/k $ implies that the longitudinal field component maintains a constant phase shift of $ \mp \pi /2 $ relative to the transversal field component. Consequently, the waveguide contains regions of elliptical polarization. Figure \ref{F1}(b) depicts the cross-sectional maps of the transverse component $ E_x $ and longitudinal component $ E_z $ multiplied with the imaginary unit. The regions with $ i{E_z}/{E_x} =  \pm 1 $, corresponding to the local circular polarization, create chirality, which results in a polarization-to-path conversion.

The chiral interaction between the quantum emitter and waveguide modes is quantified by the directional $ \beta $ factor, $ {\beta _{L/R}} = {\gamma _{L/R}}/({\gamma _L} + {\gamma _R} + \Gamma ) $, whereas the chiral contrast is defined as $ C = ({\gamma _L} - {\gamma _R})/({\gamma _L} + {\gamma _R})$, where $ {{\gamma _L}} $ ($ {{\gamma _R}} $) and $ \Gamma $ denotes the emission rate of photons into the left (right) guided mode and all other modes, respectively. By replacing $ {{\gamma _L}} $ ($ {{\gamma _R}} $) into $ {{I_{{\sigma ^ + }}}} $ ($ {{I_{{\sigma ^ - }}}} $), the chiral contrast can also be defined as $ C = ({I_{{\sigma ^ + }}} - {I_{{\sigma ^ - }}})/({I_{{\sigma ^ + }}} + {I_{{\sigma ^ - }}})$ in the experiment, where $ {{I_{{\sigma ^ + }}}} $ ($ {{I_{{\sigma ^ - }}}} $) is the photoluminescence (PL) intensity measured for the left- (right-) hand circular polarization. Figure \ref{F1}(c) shows the calculated results of the directional $ \beta $ factor and chiral contrast as a function of the displacement $ +\Delta x $ of a left-handed circularly polarized dipole ($ {\sigma ^ + } $). These factors are highly dependent on the dipole’s position. The cyan area indicates the region with a high contrast exceeding 0.85. At the position where $ {\beta _{R}} $ is close to zero and $ {\beta _{L}} $ reaches its maximum, the contrast value reaches around 1, i.e. the left-handed circularly polarized light is entirely unidirectionally transmitted to the left. The propagation direction can be changed by switching the handedness of the circularly polarized dipole. Additionally, the relative phase difference can give the handedness of the circular polarization flipping across the waveguide center; hence the chiral contrast has an antisymmetrical nature of $ C( + \Delta x) =  - C( - \Delta x) $ \cite{mrowinski2019}. Accordingly, the propagation direction can also be changed by shifting the position of the dipole along the $-x$ direction without switching the handedness of the dipole source.

A fabricated waveguide is shown in the inset of Figure \ref{F1}(d). Semicircular out-coupling gratings (OCs) with the grating period $ \lambda /2n $ were added on both sides of the waveguide for an efficient out-coupling of light. A magnetic field in the Faraday configuration which matches the QD growth direction ($y$ axis) is applied on the QDs, leading to the Zeeman splitting of $ {\sigma ^ \pm } $ polarized transitions [see the inset of Figure \ref{F1}(a) for the QD-level structure]. The QDs were non-resonantly excited and the photons were collected from the OCs. Figure \ref{F1}(d) presents the micro-photoluminescence ($ \mu $PL) spectra of an off-center QD, which has high chirality. The PL emission from the $ {\sigma ^ + }$ and $ {\sigma ^ - }$ polarized transitions were observed from the left and right OCs, respectively, with a magnetic field varying from -7 T to 7 T. Notably, regardless of the polarity of the magnetic field, the $ {\sigma ^ + }$ polarized transition always propagated to the left and $ {\sigma ^ - }$ propagated to the right. It is clearly confirmed that directional emission does not rely on emitted photon energy in this regime. The intensities of the collected opposite circularly polarized light from each OC were used to evaluate the chiral contrast. $ {C_{L}} $ was approximately $ 0.87\pm0.03$ for the left propagation mode, and $ {C_{R}} $ was approximately $-0.85\pm0.04$ for the right propagation mode, which exhibit a high directionality for this particular QD. A slight asymmetry in contrast may be attributed to the asymmetry in the quantum dot itself or the fabricated two output couplers \cite{coles2016}.

\begin{figure}
\centering
\includegraphics[width=0.95\textwidth]{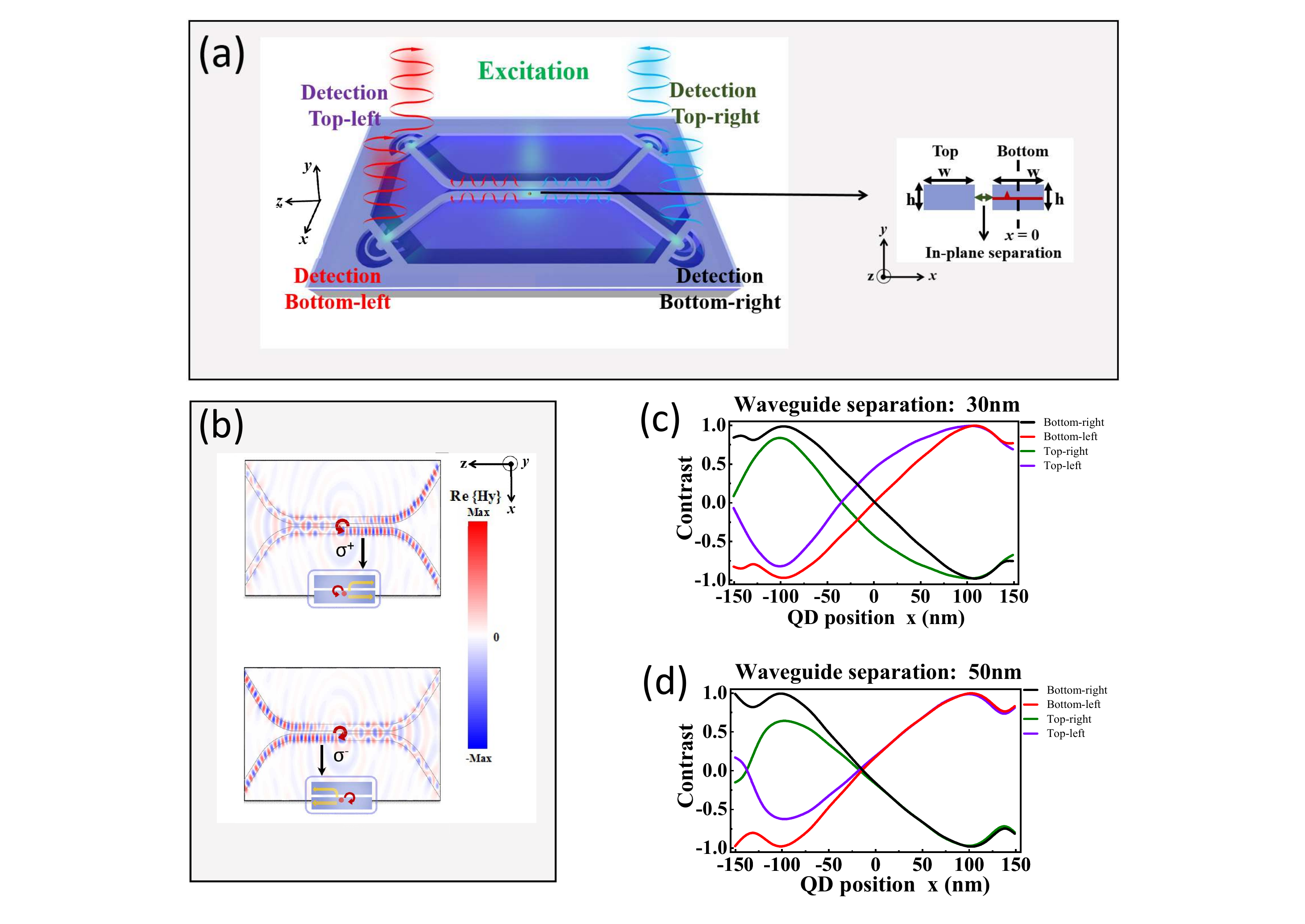}
\caption{{Design of chiral photonic circuits.} a) Schematic view of chiral photonic circuits. The inset shows the cross-sectional view ($z=0$) of the two waveguides with a QD embedded in the bottom waveguide. The PL spectra from the QD are obtained from four OCs via evanescent coupling. b) Magnetic field distributions excited by the left-handed (upper panel) and right-handed (lower panel) circularly polarized dipole located approximately $-80$ nm away from the center of the bottom waveguide. The waveguide in-plane separation is set to 20 nm, and the coupling length is set to 6 $ \mu $m. c) and d) Calculated chiral contrast of $ {\sigma ^ \pm } $ polarized light to four OCs with an in-plane separation of 30 nm and 50 nm. The source is located at a distance $x$ along the transverse direction from the bottom waveguide center to the two edges. Purple (green) curve indicates Top-left (Top-right) collection. Red (black) curve shows Bottom-left (Bottom-right) collection. }
\label{F2}
\end{figure}

\subsection{Design of chiral photonic circuits}
 Based on the chiral behavior of the single waveguide, we move to the compact chiral optical circuits. A schematic view of our chiral photonic circuits is shown in Figure \ref{F2}(a). It consists of two identical bent strip waveguides with an air gap and four OCs. The coupling of the quantum emitters takes place in the coupling region, and the emission can be transmitted and detected vertically from the four OCs. The chiral photon-emitter coupling in this structure is studied using the finite-difference time-domain method. A circularly polarized dipole is offset from the center of the bottom waveguide, as shown in the inset of Figure \ref{F2}(a). The simulated magnetic field distributions in the $ xz $-cross-section through the beam splitter for a right- and left-handed circularly polarized dipole are shown in Figure \ref{F2}(b). When the system is illuminated by a left-handed circularly polarized dipole, most of the magnetic field chirally couples to the right side and then splits into the right two OCs via evanescent field coupling (upper panel). By switching the handedness of the circularly polarized dipole, the emission direction reverses because of the time-reversal symmetry (lower panel).

We then calculate the chiral contrast of the coupled $ {\sigma ^ \pm } $ polarized light to all four OCs along the transverse direction of the waveguide. The circularly polarized dipole is set in the $ xy $ plane at $ z=0 $ (middle of the coupling region) of the bottom waveguide, as depicted schematically in the inset of Figure \ref{F2}(a). Only the coupling to the fundamental mode is considered. Figure \ref{F2}(c) and \ref{F2}(d) show the numerical calculations with two cases in which the in-plane separations of the waveguides are 30 nm and 50 nm, respectively. When the dipole is moved along the $ \pm x $ direction of the bottom waveguide, the light propagation direction changes and a strong variation in the chiral contrast is observed in all the four OCs. If the dipole is placed at a chiral point in one waveguide, where the electric field is circularly polarized, it only excites the propagation mode towards one of the two possible directions, and is then evanescently coupled to the second waveguide, i.e. the determined circularly polarized photons travel along the same side in both waveguides. Additionally, the power transferred into each OC could vary with different in-plane separations; hence, the contrast values calculated are different.

\begin{figure}
\centering\includegraphics[width=0.6\textwidth]{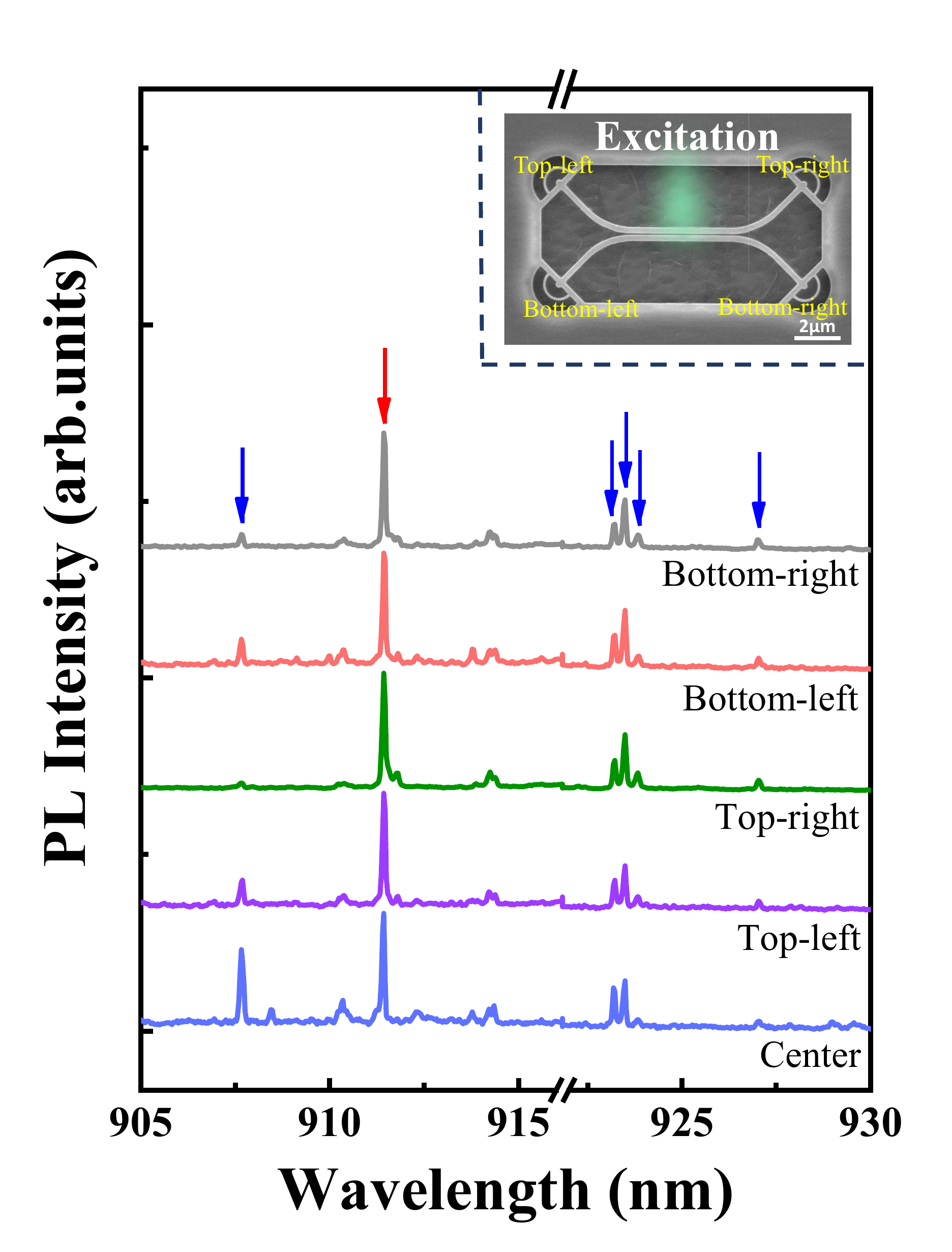}
\caption{PL spectra collected from above the coupling region and the four OCs under zero magnetic field with the excitation laser focused on the center. The arrows mark the emission lines with high transmission at four OCs, and the red arrow marks the peak with high intensity and high chiral contrast. Inset: SEM image of the chiral photonic circuit. The separation is designed with a gap of 30 nm. The QDs are embedded in the 6 $ \mu $m long coupling region. }
\label{F3}
\end{figure}

\subsection{Chiral routing and beamsplitting in photonic circuits}
Next, we experimentally explore the chiral coupling outputs from photonic circuits based on the chiral photonic device, beam splitter. The devices have a coupling length of 6 to 8 $ \mu $m and the separations between the waveguides vary from 20 to 50 nm. A scanning electron microscope (SEM) image of an evaluated chiral photonic circuit is shown in the inset of Figure \ref{F3}. We excited the QDs in the middle of the coupling region and collected emitted light from above the QDs and each OC separately. Several spectral lines operating in the 905-930 nm can be clearly observed from all four OCs (as marked in Figure \ref{F3}), demonstrating that the device has excellent transmission performance.

\begin{figure}
\centering
\includegraphics[width=\linewidth]{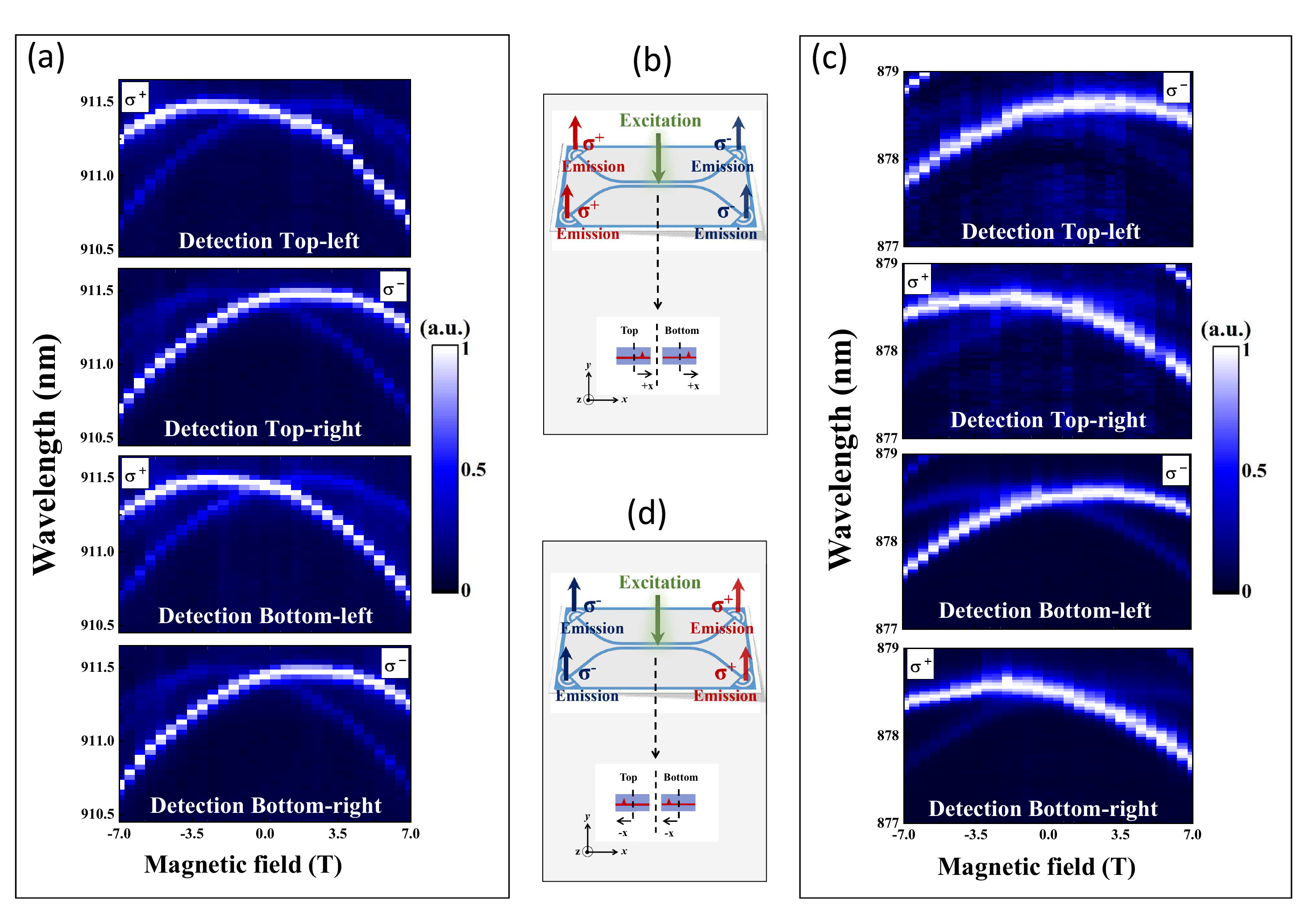}
\caption{Experimental results of the chiral behavior with deterministic spin transfer for different photonic circuits. a) Magneto-PL mappings of a QD emission (marked with red arrow in Figure \ref{F3}) detected correspondingly from the four OCs. The $ {\sigma ^ + } $ polarized light are mainly detected from both the Tl and Bl, in contrast with the $ {\sigma ^ - } $ from both the Tr and Br. b) Schematic view of spin-directionality with $ {\sigma ^ + }$ from the left two OCs and $ {\sigma ^ - }$ from the right two OCs. The QD is positioned in the $+x$ direction off-center of one of the waveguides. The vertical dashed lines in the middle of the waveguides in the cross-sectional diagram indicate the position $x = 0$ for each waveguide. c) Magneto-PL mappings of QD emission observed correspondingly from the four OCs in another device, which are opposite from the results in a). $ {\sigma ^ + } $ ($ {\sigma ^ - } $) Zeeman components are observed from Tr and Br ( Tl and Bl). d) Schematic view of spin-directionality with $ {\sigma ^ - }$ from the left two OCs and $ {\sigma ^ + }$ from the left two OCs. The QD is positioned in the $-x$ direction off-center of one of the waveguides.  }
\label{F4}
\end{figure}

Here, we focus on the strongest emission line at a wavelength of 911.4 nm (red arrow in Figure \ref{F3}). Figure \ref{F4}(a) shows the magneto-PL mappings from the four OCs, in which an asymmetry intensity for $ {\sigma ^ + }$ and $ {\sigma ^ - }$  polarized light can be observed. When detecting from the top-left (Tl) and bottom-left (Bl) OCs, strong emission from the $ {\sigma ^ + }$ branch is observed but the $ {\sigma ^ - }$ branch is weak. When detecting from the top-right (Tr) and bottom-right (Br) OCs, the asymmetric intensity was reversed corresponding to the majority of the signal originating from the $ {\sigma ^ - }$ branch. Even if the polarity of the magnetic field changed, the chiral coupling direction of the circularly polarized light was unaffected; rather, the energy of the emitted circularly polarized light was changed. From another perspective, photons with the same energy but different spin polarizations can be selected to enter the opposite OC by reversing the polarity of the magnetic field. The chiral contrasts on the four OCs can be extracted from the polarized PL intensities. For the top OCs, $ C_{Tl} $ ($ C_{Tr}$) is approximately $0.63\pm0.06$ ($-0.62\pm0.04$), and for the bottom OCs, $ C_{Bl}$ ($ C_{Br}$) is approximately $0.65\pm0.03$ ($-0.69\pm0.04$). Owing to the circular polarization distribution of the waveguide mode, we can infer that the QD was positioned probably in the $+x$ direction off-center of one of the two waveguides, as sketched in Figure \ref{F4}(b).

\subsection{Effect of QD position on out-put paths in chiral photonic circuits}
The change in the direction of the out-coupled light due to the different positions of QDs could not be observed in a single chiral photonic device because the QD is fixed in one of the waveguides. To better understand the position-dependent directional emission of single photons, we further carried out similar PL measurements on another chiral photonic device. PL spectra without a magnetic field are presented in the supporting information. The magneto-PL mappings of one of the emission lines are shown in Figure \ref{F4}(c). It is evident that $ {\sigma ^ + }$ polarized transitions were predominantly observed from the two right OCs (Tr and Br), whereas $ {\sigma ^ - }$ polarized transitions were predominantly observed from the left ones (Tl and Bl). The chiral coupling direction of the circularly polarized light of this device is exactly opposite to that in Figure \ref{F4}(a), manifesting as a reversal of the asymmetry of the emission. For the top OCs, $ C_{Tl} $ ($ C_{Tr}$) is approximately $-0.84\pm0.04$ ($0.76\pm0.05$), and for the bottom OCs, $ C_{Bl}$ ($ C_{Br}$) is approximately $0.75\pm0.06$ ($-0.72\pm0.06$). The QD was inferred to be located in the $-x$ direction off-center of one of the two waveguides, as shown in Figure \ref{F4}(d). The position-dependent chiral effect has also been affirmed in the simulation (see Figure \ref{F2}), which shows that the chiral contrast can be both positive or negative, corresponding to the change in the direction of the out-coupled light.

\begin{figure}
\centering
\includegraphics[width=0.75\textwidth]{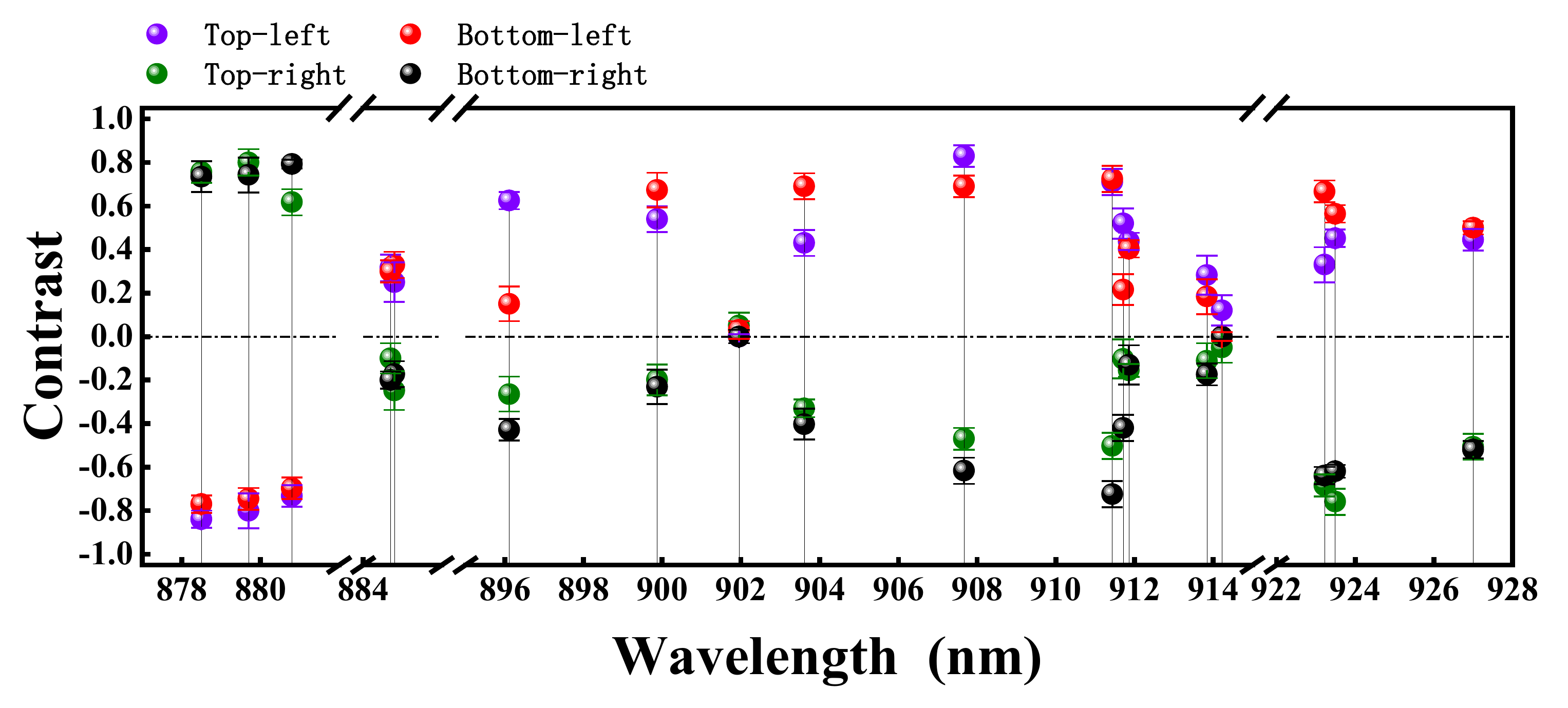}
\caption{Measured chiral contrasts of 20 emission lines from single QDs at different wavelengths from three different photonic circuits. Black straight lines correspond to the wavelength of the emission lines at zero magnetic field. Different colored dots show the chiral contrasts from different OCs. The highest chiral contrast is approximately 0.84. It should be noted that three cases in the short wavelength side are not from same device. }
\label{F5}
\end{figure}

To evaluate the chiral properties statistically, we study the spin-directionality of 20 excitonic states from single QDs in three different fabricated photonic circuits within the wavelength range of 878 nm to 928 nm at a zero magnetic field. Under a magnetic field from -7 T to 7 T, different positive or negative chiral contrast can be extracted from four OCs, as shown in Figure \ref{F5}. For most of the excitonic states, it can be observed that two left OCs have the positive chiral contrast, implying that $ {\sigma ^ + }$ polarized transitions propagate mainly to the left. Only three emission lines at short wavelengths show a chirality propagating in the opposite direction. The chiral contrasts of all the excitonic states distribute randomly, indicating that the positions of the quantum dots associated with these emission lines are randomly distributed on both sides of the center of the waveguide. It should be noted that the contrast is wavelength-insensitive owing to the weak constraint of the electric field mode within the waveguide, but is related to the handedness and purity of the circularly polarized light and position of the emitters. In our photonic structures, the wavelength of the propagating waves only affects the splitting ratio (see Supporting Information). Thus, the QDs can experience almost completely unidirectional emission in the desired direction over a broadband spectral range by choosing suitable embedding positions.

The measured chiral contrast in the chiral photonic circuits of up to 0.84 is less than the calculated result. This can be ascribed to the position of the QD, which is not at the perfect chiral point. Additionally, the back reflections of OCs and fabrication imperfections of the devices may also affect the chiral contrast. More results of the non-chirally coupled QDs are shown in Supporting Information. Nevertheless, the position-dependent directionality enables a new form of polarization control, which makes it possible to convert the handedness of the circular polarization on the same waveguide OC for more functionality. More importantly, the beam splitter structure provides a novel spin-photon interface in the chiral photonic circuits, with an ability of chiral routing and beamsplitting of single photons in a deterministic way \cite{lodahl2017}.

\section{\label{sec3}Conclusion}

In conclusion, we demonstrated compact chiral photonic circuits with a quantum light source on-chip for deterministic spin transfer. The spin-momentum locking induced by the intrinsic transverse spin inside the waveguide and interaction of the guided mode via the evanescent field between two adjacent waveguides provide the fundamental schemes of routing and splitting chiral photons. The position-dependent directional emission of the QD was demonstrated in the chiral photonic circuits, enabling the control of the in-plane transfer direction of the spin states with a high chiral contrast of up to 0.84. With the QD position registered precisely in the future, such photonic circuits with specified spin outputs could be fabricated in a deterministic way, and the directionality and efficiency of the photonic chiral behaviors could be substantially improved, to reduce crosstalk and insertion loss in the circuits. In addition, dynamically adjusting the arbitrary proportional beam splitting of the circularly polarized light could also be implemented using the in-plane or out-of-plane electro-mechanical actuation of a cantilever \cite{bishop2018,papon2019nanomechanical}. The chiral photonic circuits with chiral light-matter interaction provides a base for multi-functional devices, such as quantum spin interference \cite{tang2015quantum}, optical logic circuit \cite{pittman2001}, quantum information  storage \cite{maddox1987quantum,sillanpaa2007coherent}, and muli-photon entanglement \cite{lim2005,pan2012multiphoton}, which makes the chiral quantum photonic network on-chip feasible in the future.

%

\section{\label{sec4}Experimental Section}

\subsection{Sample Fabrication}
The samples were grown by molecular beam epitaxy and consisted of a $ 1~\mu$m AlGaAs sacrificial layer on a GaAs substrate. On top of this, a 150 nm GaAs membrane was deposited with a single layer of self-assembled InGaAs QDs embedded at the center. The samples were first spin-coated with a positive resist AR-P 6200. The patterns of the nanophotonic structures were then defined by the electron beam lithography technique. The exposed regions were then developed using a developer AR 600-546, leaving a mask for etching processes. Inductively coupled plasma (ICP) etching with gases of $ \rm BCl_{3} $ and Ar was used to transfer the patterns to the GaAs membrane. Finally, the AlGaAs sacrificial layer was removed by selective wet etching using the HF solution to leave free a suspended GaAs membrane containing the designed structures.

\subsection{Experimental Setup}
 All the experiments were performed in a helium bath cryostat. The device was mounted on a three-axis piezoelectric positioner and cooled down to 4.2 K via heat change with helium gas. The superconducting magnet surrounding the device provided a Faraday-geometry magnetic field of up to 9 T. Two independently-controlled optical paths were used for exciting the QDs and collecting the PL spectra from different locations. The surface of the device was imaged by a light-emitting diode illumination and a complementary metal oxide semiconductor camera, allowing us to inspect the position of the device and locate the excitation and collection spots. Spatially-selective excitation and collection were through a confocal $ \mu $PL setup with a 0.8 numerical aperture objective lens. The spot size was approximately 1 $ \mu $m in diameter. The QDs were non-resonantly excited by a continuous-wave laser with a wavelength of 532 nm and the PL emission from the OC could be focused on an optical fiber to achieve spatial filtering. The collected signal was then dispersed by a single 0.55 m spectrometer and recorded with a liquid nitrogen cooled charged coupled device camera with a resolution of 60 $ \mu $eV.

\section{\label{sec5}Acknowledgments}

This work was supported by the National Natural Science Foundation of China (Grants No. 62025507, No. 11934019, No. 11721404, and No. 11874419), the Key-Area Research and Development Program of Guangdong Province (Grant No.2018B030329001), and the Strategic Priority Research Program (Grant No. XDB28000000), the Instrument Developing Project (Grant No. YJKYYQ20180036) and the Interdisciplinary Innovation Team of the Chinese Academy of Science.

\end{document}